\documentstyle[prl,aps,twocolumn,latexsym,amsfonts,epsf,graphicx]{revtex}

\newcommand{\nn}{\nonumber\\}

\newcommand{\bea}{\begin{eqnarray}}
\newcommand{\ea}{\end{eqnarray}}
\newcommand{\eea}{\end{eqnarray}}

\begin{document}

\wideabs{
\title{A relativistic toy model for back-reaction}
\author{G\"unter Plunien, Marcus Ruser, and Ralf Sch\"utzhold}
\address{Institut f\"ur Theoretische Physik,
Technische Universit\"at Dresden,
01062 Dresden, Germany}
\date{\today}
\maketitle
\begin{abstract} 
We consider a quantized massless and minimally coupled scalar field 
on a circular closed string with a time-dependent radius $R(t)$, whose
undisturbed dynamics is governed by the Nambu-Goto action. 
Within the semi-classical treatment, the back-reaction of the quantum
field onto the string dynamics is taken into account in terms of the
renormalized expectation value of the energy-momentum tensor including
the trace anomaly.   
The results indicate that the back-reaction could prevent the collapse
of the circle $R\downarrow0$ -- however, the semi-classical picture
fails to describe the string dynamics at the turning point  
(i.e., possible bounce) at finite values of $R$ and $\dot R$.
The fate of the closed string after that point (e.g., oscillation or
eternal acceleration) cannot be determined within the semi-classical
picture and thus probably requires the full quantum treatment.
\end{abstract} 
}

\section{Introduction}

Without knowledge of the underlying theory including quantum gravity
the back-reaction of quantum fields onto the geometry is usually
described by the semi-classical Einstein equations 
\bea
\label{einstein}
{\cal R}_{\mu\nu}-\frac12{g}_{\mu\nu}\,{\cal R}
=-\kappa\,\langle\hat T_{\mu\nu}\rangle_{\rm ren}
\,.
\ea
Although intrinsically incomplete, this semi-classical approach might 
-- after proper renormalization of 
$\langle\hat T_{\mu\nu}\rangle_{\rm ren}$ is employed -- 
shed light onto the following questions:
%
\begin{itemize}
\item
Could the back-reaction of quantum fields prevent the collapse 
(i.e., singularity) of the space-time? 
\item
Under which circumstances does the semi-classical picture apply and
when does it fail? 
\end{itemize}

As a model for gauge field theories (vector $A_\mu$ and spinor $\Psi$ fields) 
in a 3+1 dimensional gravitational background we shall consider the
also conformally invariant theory of a massless and minimally coupled
scalar field $\Phi$ in 1+1 dimensions.
Since the Einstein tensor ${\cal R}_{\mu\nu}-{g}_{\mu\nu}\,{\cal R}/2$
vanishes identically in 1+1 dimensions,  
one has to start with an alternative geometric action leading to a
preferably simple dynamics which still preserves major aspects such
as relativistic invariance as well as generic features of
higher-dimensional space-times.

\section{Toy model}

Let us consider a 1+1 dimensional string whose dynamics is governed by
the Nambu-Goto action \cite{nambu}
\bea
\label{string}
{\mathfrak A}_{\rm string}=-\frac{\sigma}{2}\int d^2x\,\sqrt{-{g}}
\,,
\ea
with $\sigma$ denoting the string tension.
For simplicity we shall assume rotational symmetry, i.e., a circular
closed string characterized by its time-dependent radius $R$ as the
only degree of freedom of the geometry.
Adopting the usual time-gauge (see, e.g. \cite{nambu,vilenkin}), 
the induced metric reads ($\hbar=c=1$ throughout) 
\bea
\label{time-gauge}
ds^2=\left[1-(\partial_tR)^2\right]\,dt^2-R^2d\varphi^2
\,,
\ea
where $t$ is the ''laboratory'' time and $\varphi$ the angular
coordinate. 
As one can immediately see, the action in Eq.~(\ref{string}) is fully
relativistic (in contrast to any {\em ad hoc} kinetic terms like 
$\dot R^2=(\partial_t R)^2$).

Left alone -- i.e., without additional forces -- the free dynamics of
the circle governed by $\delta{\mathfrak A}_{\rm string}/\delta R=0$,
i.e.,
\bea
R\ddot{R}+(1-\dot{R}^2)=0
\label{free EOM}
\,,
\ea 
with the initial conditions $R(0)=R_0$, $\dot{R}(0)=0$, for example, 
leads to a collapse $R\downarrow0$ and $\dot R\uparrow1$ 
(i.e., a singularity) 
after a finite time caused by the string tension $\sigma>0$.

Let us consider this circle (time-dependent background metric) being
endowed with a real scalar field $\Phi$ such that the total action
reads (see, e.g., \cite{diplom})
\bea
\label{total-action}
{\mathfrak A}=\frac{1}{2}\int d^2x\,\sqrt{-{g}}
\left(\partial_\mu\Phi\,\partial^\mu\Phi-\sigma\right)
\,.
\ea
The (classical) equations of motion for $R(t)$ can be derived by
variation $\delta{\mathfrak A}/\delta R=0$
\begin{equation}
\label{eom}
\sqrt{1-\dot{R}^2}\left[\frac{\sigma}{2}+T^{1}_{1}
\right]+\frac{d}{dt}\left[\frac{R\dot{R}}{\sqrt{1-\dot{R}^2}}
\left(\frac{\sigma}{2}+T^{0}_{0}\right)\right]=0
\,,
\end{equation}
with the energy-momentum tensor of the scalar field 
\bea
\label{Tmunu}
T_{\mu\nu}=\partial_\mu\Phi\,\partial_\nu\Phi-\frac12\,
{g}_{\mu\nu}\partial_\rho\Phi\,\partial^\rho\Phi
\,.
\ea
The equation of motion for scalar field 
$\delta{\mathfrak A}/\delta\Phi=0$ 
follows as
\bea
\Box\Phi=\frac{1}{\sqrt{-g}}\,\partial_\mu\left(
\sqrt{-g}\,g^{\mu\nu}\,\partial_\nu\,\Phi\right)=0
\, ,
\ea
where $\Box$ denotes the d'Alembertian with respect to the metric in 
Eq.~(\ref{time-gauge}).

The first integral of the equations of motion (\ref{eom}) and
$\Box\Phi=0$ corresponds to the (conserved) total energy
\begin{eqnarray}
\label{energy}
E=\frac{2\pi\,R}{\sqrt{1-\dot{R}^2}}\left[\frac{\sigma}{2}+
T^{0}_{0}\right] 
\,.
\end{eqnarray}
%

\section{Classical case}

For calculating the dynamics of the field $\Phi$ it is convenient to
introduce another time coordinate via the transformation
\bea
\label{trafo}
\tau=\int dt\,\frac{\sqrt{1-(\partial_tR)^2}}{R(t)}
\,,
\ea
leading to conformally flat metric
\bea
\label{conformal}
ds^2=R^2(\tau)[d\tau^2-d\varphi^2]
\,.
\ea
Note that the insertion of this metric into the Nambu-Goto action in
Eq.~(\ref{string}) does not yield the same equations of motion for
$R(t)$ since Eq.~(\ref{trafo}) represents a non-algebraic transformation.

Let us first consider the classical back-reaction of a spatially
homogeneous field $\partial\Phi/\partial\varphi=0$ as a solution of
the field equation $\Box\Phi=\partial^2\Phi/\partial\tau^2=0$ which
simply reads $\Phi={\mathfrak C}_\Phi\tau$ leading to the kinetic
term 
\bea
\dot\Phi^2={\mathfrak C}_\Phi^2\,
\frac{1-\dot R^2}{R^2}
\,,
\ea
with respect to the time $t$.
Insertion into the total energy in Eq.~(\ref{energy}) yields
\begin{eqnarray}
\label{class energy}
E=\frac{2\pi\,R}{\sqrt{1-\dot{R}^2}}\left[\frac{\sigma}{2}+
\frac12\,\frac{{\mathfrak C}_\Phi^2}{R^2}
\right] 
\,.
\end{eqnarray}
Since this energy becomes arbitrarily large for both, $R\downarrow0$
and $R\uparrow\infty$ (as well as for $\dot{R}^2\uparrow1$) the
presence of this (classical) scalar field solution prevents the
collapse $R\downarrow0$ for ${\mathfrak C}_\Phi\neq0$.
The dynamics is governed by 
\bea
R\ddot{R}\left[R^2+R_E^2\right]+(1-\dot{R}^2)
\left[R^2-R_E^2 \right]=0
\,,
\ea
with $R_E^2={\mathfrak C}_\Phi^2/\sigma>0$ denoting the square of the 
radius which minimizes the energy in Eq.~(\ref{class energy}) for
$\dot{R}=0$. 
Initial conditions $R(t=0)\neq R_E$ or $\dot{R}(t=0)\neq0$ lead to an
eternal oscillation between $R_{\rm min}>0$ and $R_{\rm max}$ with
$R_{\rm min}<R_E<R_{\rm max}$ whereas for $R(t=0)=R_E$ and
$\dot{R}(t=0)=0$, the ring remains static forever.  

However, it should be emphasized here that a spatially homogeneous
solution such as $\partial\Phi/\partial\varphi=0$ -- which is also 
called a zero-mode -- does usually not exist in higher-dimensional 
situations with vector fields, for example (see also the next Section). 
On the other hand, one would obtain the same result for a thermal
bath of the $\Phi$-field with $\langle E \rangle_\beta$ representing
the (classical) ensemble average of the energy for the inverse
temperature $\beta$.
The calculation of $\langle T^0_0 \rangle_\beta$ may proceed in
basically the same way as the derivation starting from
Eq.~(\ref{balance}) below, provided that one neglects all quantum
effects such as the trace anomaly. 

As we shall see in the next Section, the Casimir effect contributes to
the total energy in the same way, but with a negative sign. 
Therefore, the above (classical) contribution and the induced
stabilization of the string can be negated by the (quantum) Casimir
effect, in particular since the amplitude of  the field or the
temperature can decrease -- whereas the Casimir effect remains. 
This observation motivates the consideration of the quantum field
effects. 

\section{Renormalization of $\langle\hat T_{\mu\nu}\rangle$ and trace anomaly}

In order to investigate the back-reaction of the quantum field
$\hat\Phi$ in analogy to the semi-classical Einstein equations
(\ref{einstein}), we have to calculate the renormalized expectation
value of the energy-momentum tensor and insert it into the equation of
motion (\ref{eom}) for $R(t)$.
In both cases, the renormalization of $\langle\hat T_{\mu\nu}\rangle$,
i.e., the absorption of the divergences, requires a redefinition of
the involved coupling constants -- the cosmological constant $\Lambda$
and Newton's constant $\kappa$ in gravity, and the string tension
$\sigma$ in our model.

Adopting the point-splitting renormalization procedure, we need the 
two-point function $W$.  
Fortunately, a massless and minimally coupled scalar field in 1+1
dimensions is conformally invariant and hence the two-point function
of the conformal vacuum (see, e.g., \cite{birrell}) in terms of the
conformal coordinates in Eq.~(\ref{conformal}) has the same for as in
flat space-time ($R=\rm const$)
\bea
\label{wightman}
\langle\hat\Phi(\tau,\varphi)\hat\Phi(\tau',\varphi')\rangle=
W(\tau-\tau',\varphi-\varphi')
\,.
\ea
In the coincidence limit, i.e., for $\tau\to\tau'$ and 
$\varphi\to\varphi'$, the Wightman function $W$ behaves as 
$\ln[(\tau-\tau')^2-(\varphi-\varphi')^2]$.

Note that $W$ is only determined up to an additive constant -- 
in the usual 1+1 dimensional space-time ${\mathbb R}\times{\mathbb R}$, 
this fact reflects the infra-red problem (in 1+1 dimensions); and, in
our case ${\mathbb R}\times{\mathbb S}_1$, this undetermined additive
constant corresponds to the zero mode $\partial\Phi/\partial\varphi=0$.

However, zero modes do not have a zero-point energy 
(the energy spectrum reaches zero) and decouple from the rest of the
modes (in our situation). 
Therefore, we omit those modes and the related problems with their
quantization in the following.

In any covariant regularization method in 1+1 dimensions
(where only one principal divergence exists), 
for example point splitting after omitting direction dependent terms,
the regularized (but not yet renormalized) expectation value reads
(see also Appendix A)
\bea
\label{reg-T}
\langle\hat T_{\mu\nu}\rangle_{\rm bare}
=
\frac{1}{\epsilon}\,{g}_{\mu\nu}+
\langle\hat T_{\mu\nu}\rangle_{\rm ren}
\,,
\ea
where $\epsilon\to0$ is the regularization parameter 
(e.g., the geodesic distance in point-splitting) and the remaining 
$\langle\hat T_{\mu\nu}\rangle_{\rm ren}$ is finite.

In the expressions for the energy (\ref{energy}) as well as the
equations of motion (\ref{eom}), the above divergence can be
completely absorbed by a renormalization of string tension via
\bea
\label{reg-sigma}
\sigma_{\rm bare}
=
-\frac{2}{\epsilon}+
\sigma_{\rm ren}
\,.
\ea
The question of whether $\sigma_{\rm ren}$ can acquire any dependence
on $R$ or $\dot R$ via the renormalization procedure (roughly similar
to the running coupling in QED, for example) will be answered below.

Owing to the conformal invariance of the scalar field, the two-point 
function of the conformal vacuum can be calculated easily in the 
conformal metric, for example for a single scalar field with periodic
boundary conditions we obtain
\bea
\label{periodic}
&&
\langle\partial_\varphi\hat\Phi(\tau,\varphi)
\partial_{\varphi'}\hat\Phi(\tau,\varphi')\rangle
=
\langle\partial_\varphi\hat\Phi(t,\varphi)
\partial_{\varphi'}\hat\Phi(t,\varphi')\rangle
=
\nn
&&
=
\frac{1}{4\pi}\,\frac{1}{\cos(\varphi-\varphi')-1}
\,.
\ea
By means of this example, one can read off the symmetries 
$\varphi\to\varphi+\varphi_0$ and $\varphi\to-\varphi$
which reflect the homogeneity and ${\mathbb Z}_2$-isotropy of the
underlying space-time.
These symmetries of the vacuum state and the geometry are inherited by
the (renormalized) expectation value of the energy-momentum tensor
(see, e.g., \cite{trace,birrell})
\bea
\label{flux}
\langle\hat T_0^1\rangle_{\rm ren}=\langle\hat T_1^0\rangle_{\rm ren}=0
\,,
\ea
i.e., no flux, as well as
\bea
\label{homo}
\partial_\varphi\langle\hat T^\mu_\nu\rangle_{\rm ren}=0
\,.
\ea

Furthermore, one has to demand that the renormalization procedure of
$\langle\hat T^\mu_\nu\rangle_{\rm ren}$ respects the property of a
vanishing covariant divergence (see also Appendix A)
\bea
\label{balance}
\nabla_\mu\langle\hat T^\mu_\nu\rangle_{\rm ren}
&=&0\qquad\leadsto
\nn
\partial_\mu\left(\sqrt{-{g}}\,
\langle\hat T^\mu_\nu\rangle_{\rm ren}\right)
&=&
\frac{\sqrt{-{g}}}{2}\,\langle\hat T^{\alpha\beta}\rangle_{\rm ren}
\partial_\nu {g}_{\alpha\beta}
\,.
\ea
As it is well known \cite{trace}, one has to give up the classical
feature $T^\rho_\rho=0$ in this process, i.e., 
$\langle\hat T^\mu_\nu\rangle_{\rm ren}$ acquires an anomalous trace
during the renormalization 
\bea
\label{anomaly}
\langle\hat T_\mu^\mu\rangle_{\rm ren}={\mathfrak C}_{\rm tr}\,{\cal R}
\,,
\ea
with ${\mathfrak C}_{\rm tr}$ denoting a constant related to the
central charge. 
In 1+1 dimensions, the trace anomaly is completely determined by the
Ricci scalar (see, e.g., \cite{trace,birrell})
\bea
\label{ricci}
{\cal R}
=
2\,\frac{\ddot{R}}{R(1-\dot{R}^2)^2}
=
2\,\frac{R\,R''-(R')^2}{R^4}
\,,
\ea
with $\dot{R}=dR/dt$ and $R'=dR/d\tau$, respectively.

Similar to the calculation of the Hawking radiation in 1+1 dimensions
\cite{trace} via the trace anomaly, we can integrate 
Eq.~(\ref{balance}) with the proper boundary/initial conditions and 
symmetries. 
The case $\nu=1$ in Eq.~(\ref{balance}) is trivial and in the
remaining equation with $\nu=0$ -- when expressed in terms of the
conformal metric -- the trace anomaly acts somewhat similar to a
source 
\bea
\label{source}
\partial_\tau(R^2\langle\hat T^0_0\rangle_{\rm ren})=
2\,\frac{R\,R'\,R''-(R')^3}{R^3}\,{\mathfrak C}_{\rm tr}
\,.
\ea
The conformal invariance of the scalar field in 1+1 dimensions leading
to the absence of particle creation (conformal vacuum) manifests
itself in the fact that the r.h.s.\ of the above equation is a total
differential\footnote{If particles were created, the energy density 
$\langle\hat T^0_0\rangle_{\rm ren}$ would depend on the history of
the space-time and not just on the values $R(t)$ and $\dot R(t)$ etc.,
at present.}.
Thus integrating Eq.~(\ref{source}) yields
\bea
\label{T00-conf}
\langle\hat T^0_0\rangle_{\rm ren}
={\mathfrak C}_{\rm tr}\,\frac{(R')^2}{R^4}+
{\mathfrak C}_{\rm Cas}\,\frac{1}{R^2}
\,,
\ea
where ${\mathfrak C}_{\rm Cas}$ is {\em a priori} an integration constant. 
However, if we consider a static circle $R=\rm const$, the only
contribution to the vacuum energy is the Casimir effect induced by the
compactness of the space (${\mathbb S}_1$).
In this case ${\mathfrak C}_{\rm Cas}/R^2$ is just the Casimir energy
density which allows us to determine the constant ${\mathfrak C}_{\rm Cas}$
and we keep this nomenclature also for the general time-dependent
situation $R'\neq0$. 

Transforming back to the laboratory time we obtain 
(note that $\tau=\tau(t)\,\to\,t$ and hence $T^0_0\,\to\,T^0_0$)
\begin{equation}
\label{T00-lab}
\langle\hat T^0_0\rangle_{\rm ren}
=\frac{1}{R^2}\left[
\frac{\dot{R}^2}{1-\dot{R}^2}\;{\mathfrak C}_{\rm tr}+
{\mathfrak C}_{\rm Cas}\right]
\,.
\end{equation}
The remaining non-trivial component $\langle\hat T^1_1\rangle_{\rm ren}$
can be determined via the trace anomaly
\begin{equation}
\label{T11-lab}
\langle\hat T^1_1\rangle_{\rm ren}=
2\;{\mathfrak C}_{\rm tr}\;\frac{\ddot{R}}{R(1-\dot{R}^2)^2}
-\langle\hat T^0_0\rangle_{\rm ren}
\,.
\end{equation}
%

\section{Back-reaction}

At this stage we are in the position to calculate the back-reaction in
analogy to 
the semi-classical Einstein equations by inserting the renormalized
expectation values of the energy-momentum tensor 
$T^\mu_\nu\,\to\,\langle\hat T^\mu_\nu\rangle_{\rm ren}$
in Eqs.~(\ref{T00-lab}) and (\ref{T11-lab}) into the equation of motion
(\ref{eom}) 
\begin{eqnarray}
\label{eom-ren}
R\ddot{R}\left(
\frac{\sigma_{\rm ren}}{2}R^2+{\mathfrak C}_{\rm Cas}
+
\frac{2+\dot{R}^2}{1-\dot{R}^2}\,
{\mathfrak C}_{\rm tr}
\right)
+
\nonumber\\
+
(1-\dot{R}^2)
\left(
\frac{\sigma_{\rm ren}}{2}R^2-{\mathfrak C}_{\rm Cas}
\right)
-\dot{R}^2{\mathfrak C}_{\rm tr}
=0
\,.
\end{eqnarray}
The consistency of the above equation -- assuming a  constant
$\sigma_{\rm ren}$ -- with the conservation of the total
(renormalized) energy of the system  
\begin{eqnarray}
\label{energy-ren}
E
&=&
\frac{2\pi\,R}{\sqrt{1-\dot{R}^2}}
\left(\frac{\sigma_{\rm ren}}{2}+
\langle\hat T^0_0\rangle_{\rm ren}
\right) 
=
\frac{2\pi\,R}{\sqrt{1-\dot{R}^2}}
\times
\nn
&&\times
\left(
\frac{\sigma_{\rm ren}}{2}+
\frac{1}{R^2}\left[
\frac{\dot{R}^2}{1-\dot{R}^2}\;{\mathfrak C}_{\rm tr}+{\mathfrak C}_{\rm Cas}
\right]\right)
\,,
\end{eqnarray}
can be considered as a cross-check
that $\sigma_{\rm ren}$ does not acquire any non-trivial dependence on
$R(t)$ or $\dot R(t)$ during the renormalization procedure
(see also Appendix A).

It turns out that the above equation of motion (\ref{eom-ren}) can be
derived from the following effective Lagrangian
\bea
\label{Leff}
L_{\rm eff}=2\pi\,\frac{\sqrt{1-\dot R^2}}{R}
\left(
-\frac{\sigma_{\rm ren}\,R^2}{2}
-{\mathfrak C}_{\rm Cas}+
\frac{\;{\mathfrak C}_{\rm tr}\dot{R}^2}{1-\dot{R}^2}
\right)
\,,
\ea
and that the total (renormalized) energy in Eq.~(\ref{energy-ren}) is
just the associated Hamiltonian $E=H=P_{\rm eff}\dot R-L_{\rm eff}$.

\begin{figure}[!h]
\includegraphics[height=7cm]{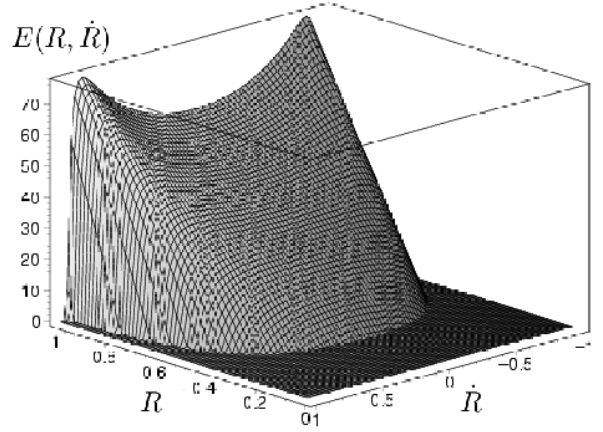}
\caption{Energy landscape (in arbitrary units) with $\sigma=20$ and  
${\mathfrak C}_{\rm tr}={\mathfrak C}_{\rm Cas}=-1$, 
which has been cut off at negative energies for convenience.
As one can easily perceive, a circle with a positive (initial) energy
can never collapse -- i.e., reach $R=0$ -- unless it leaves the region
of validity of the semi-classical picture.}
\label{full}
\end{figure}

\begin{figure}[!h]
\includegraphics[height=7cm]{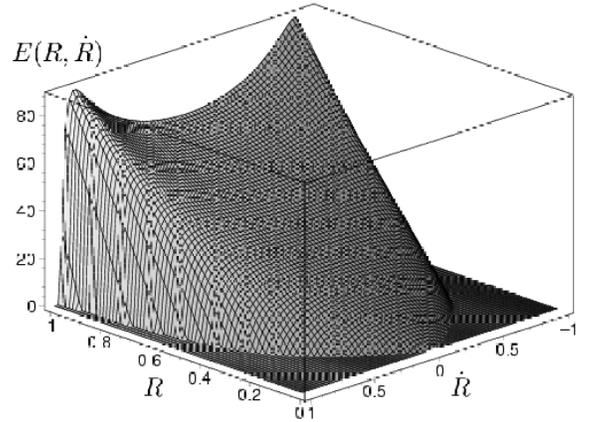}
\caption{Energy landscape with the same values for $\sigma=20$ and
${\mathfrak C}_{\rm tr}=-1$, but no Casimir effect 
${\mathfrak C}_{\rm Cas}=0$. Hence the Casimir effect is not
substantial for preventing the collapse to $R=0$.} 
\label{no-cas}
\end{figure}

As for any conservative system, possible trajectories $R(t)$ can be
inferred from the energy landscape according to
Eq.~(\ref{energy-ren}), see Fig.~\ref{full}.
To this end we have to specify the constants ${\mathfrak C}_{\rm tr}$
and ${\mathfrak C}_{\rm Cas}$.
For the example of a single scalar field with periodic boundary
conditions we have
\bea
\label{single}
{\mathfrak C}_{\rm tr}=-\frac{1}{24\pi}
\;,\;
{\mathfrak C}_{\rm Cas}=-\frac{\pi}{6}
\,.
\ea
For multiple scalar fields and/or possibly different boundary
conditions one would obtain other values.
Nevertheless, both contributions remain always negative ${\mathfrak
C}_{\rm tr}<0$ and ${\mathfrak C}_{\rm Cas}<0$. 
In this situation one can infer from the qualitative behavior of the 
energy landscape in Figs.~\ref{full} and \ref{no-cas} that a
trajectory $R(t)$ corresponding to a positive energy $E>0$ does never
reach $R=0$ and also cannot approach $\dot R=1$ for finite $R$.
I.e., we avoid the collapse to a singularity $R=0$ -- provided that
the semi-classical treatment does not break down completely
(cf.~the next Section).
%

\section{Bounce}

It turns out to be very interesting to discuss possible trajectories
$R(t)$ in more detail. 
Let us assume that we start with a circle with a large initial radius
$R(t=0)=R_0$ and a vanishing initial velocity $\dot R(t=0)=0$ -- where
the quantum effects are negligible. 
In the energy landscape in Fig.~\ref{full}, this corresponds
to the region in the middle (around the axis $\dot R=0$) high up the 
mountain. 
Starting in normal region with $R(t=0)=R_0$, the radius begins to
shrink $\dot R<0$ and the negative velocity gradually increases in
magnitude.  
After a while, i.e., for a sufficiently large velocity 
(and, correspondingly, small radius), quantum effects become
important.  
Finally, the trajectory reaches the ridge, where the radius $R$ cannot
decrease anymore.
Therefore, the velocity $\dot R$ should change its sign -- i.e., the 
trajectory $R(t)$ should turn around.
But these two ridges correspond to finite velocities $\dot R$ -- one
positive and the other negative -- such that a change in the sign is only
possible by jumping from one ridge to the other
(corresponding to the same values of $R$, $|\dot R|$, and $E$).
Obviously this requires a diverging $\ddot R$ and seems to be a very
strange property of the semi-classical system. 

Nevertheless, it turns out that such a strange property is a necessary
consequence of the weird features of the energy landscape in
Fig.~\ref{full}. 
The energy $E(R,\dot R)$ equals the Hamiltonian $H(R,P)$ derived from
the effective Lagrangian $L(R,\dot R)$ in Eq.~(\ref{Leff}).
For a given and fixed radius $R$, there are three values of $\dot R$ 
for which
\bea
\left(\frac{\partial E}{\partial\dot R}\right)_R=0
\,,
\ea
the usual $\dot R=0$ (in the middle, i.e., the normal region) and two
anomalous points (on the two ridges) where $\dot R\neq0$. 
In view of the Hamilton equation
\bea
\dot R
=
\left(\frac{\partial H}{\partial P}\right)_R
=
\left(\frac{\partial H}{\partial\dot R}\right)_R
\left(\frac{\partial\dot R}{\partial P}\right)_R
\,,
\ea
for these two anomalous points where $\dot R\neq0$, we must have 
\bea
\left(\frac{\partial P}{\partial\dot R}\right)_R
=
\left(\frac{\partial^2L}{\partial\dot R^2}\right)_R
=0
\,.
\ea
I.e., the ridges are critical points where the Euler-Lagrange equation 
\bea
\left(\frac{\partial L}{\partial R}\right)_{\dot R}
=
\frac{d}{dt}\left(\frac{\partial L}{\partial\dot R}\right)_R
=
\ddot R\left(\frac{\partial^2L}{\partial\dot R^2}\right)_R
+
\dot R\,\frac{\partial^2L}{\partial R\,\partial\dot R}
\,,
\ea
cannot determine $\ddot R$ anymore, i.e., $\ddot R$ diverges.
Indeed, the pre-factor in front of the $\ddot R$-term in the equation
of motion (\ref{eom}) goes to zero at these critical points.

Since $\ddot R$ approaches infinity there -- whereas $\dot R$ and $R$
remain finite --  the Ricci scalar diverges at the ridges, i.e., these
critical points still represent a curvature singularity.
\begin{figure}[!h]
\includegraphics[height=6.5cm]{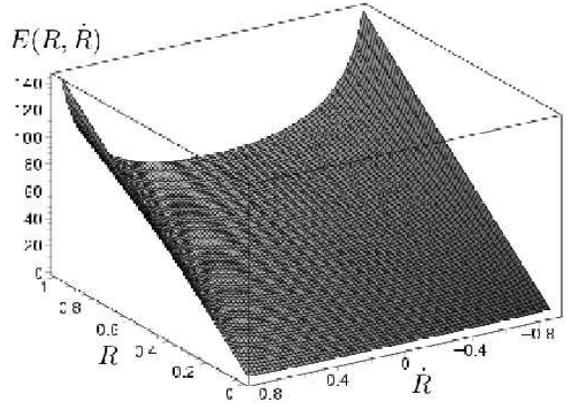}
\caption{Energy landscape for the classical case, i.e.,
${\mathfrak C}_{\rm tr}={\mathfrak C}_{\rm Cas}=0$. All trajectories
corresponding to some (positive) energy inevitably lead to
$R\downarrow0$ and $\dot R\downarrow-1$, i.e., the circle collapses.}
\label{class}
\end{figure}
\begin{figure}[!h]
\includegraphics[height=6.5cm]{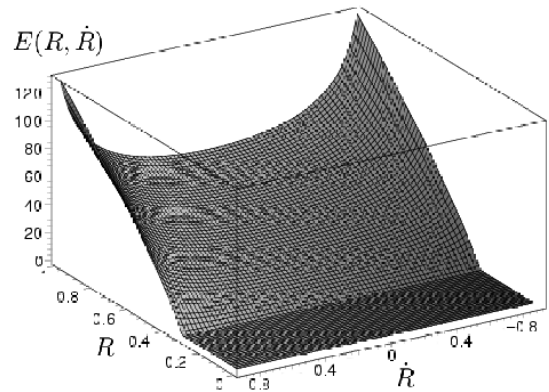}
\caption{Energy landscape for ${\mathfrak C}_{\rm tr}=0$ and 
${\mathfrak C}_{\rm Cas}=-1$. In this rather artificial set-up the
trajectories reach the boundary $|\dot R|=1$ at a finite radius $R>0$.}
\label{no-tr}
\end{figure}
%

\section{Conclusions}

Let us compare three scenarios:
\begin{itemize}
\item Without any back-reaction (cf.~Fig.~\ref{class}), the string
collapses to $R=0$ and $\dot R=1$ after a finite laboratory time --
i.e., the metric, the action and the Ricci scalar become singular.
\item With purely classical back-reaction (e.g., a thermal bath) all
of these singularities are avoided and the string oscillates smoothly
forever. However, the Casimir effect generates the same contribution
-- but with the opposite sign -- and can therefore negate this effect. 
\item If we take into account the Casimir effect only -- which is
rather artificial -- the string reaches $\dot R=1$ and hence develops
a singularity of the metric, the action and the Ricci scalar,
respectively, after a finite laboratory time and at a finite radius
$R>0$ (cf.~Fig.~\ref{no-tr}). 
\item Calculating the full expectation value 
$\langle\hat T^\mu_\nu\rangle_{\rm ren}$ in the conformal vacuum --
including the trace anomaly -- we find again that circle reaches the 
critical point after a finite laboratory time, cf.~Fig.~\ref{full}.
However, such a critical point is characterized by a singular
acceleration $\ddot R$ and hence Ricci scalar only -- while the radius
$R>0$, the velocity $\dot R<1$, and hence both, the action as well as
the metric remain regular. 
\end{itemize}
We observe that the back-reaction of the quantum field $\hat\Phi$
prevents the strong singularity of the first scenario, but still
generates a weaker singularity.
At this critical point (i.e., the ridge in the energy landscape) the
pre-factor in front of $\ddot R$ obtained via the semi-classical
treatment vanishes. 
It appears natural to expect that the quantum back-reaction beyond the 
semi-classic picture will become important in the vicinity of this point.

It is evident that the semi-classical treatment of the back-reaction
problem has clear problems, cf.~\cite{example}. 
The major point is that the semi-classical Einstein equations 
-- and similarly Eq.~(\ref{eom-ren}) --
imply the neglect of all quantum fluctuations by considering the
expectation value $\langle\hat T_{\mu\nu}\rangle_{\rm ren}$ only.
This approximation is justified if the fluctuations are sufficiently
small, i.e., as long as conditions such as 
\bea
\langle\hat T_{\mu\nu}^2\rangle_{\rm ren}
\stackrel{?}{\approx}
\langle\hat T_{\mu\nu}\rangle_{\rm ren}^2
\,,
\ea
hold.
Assuming the validity of the renormalization procedure such as
point-splitting -- e.g., that all allowed quantum states satisfy the 
Hadamard condition -- the trace-anomaly is independent of the quantum
state and hence a c-number.
Ergo, for this particular quantity one would expect the
semi-classical treatment to work.
However, the equations of motion do not only contain
$\langle\hat T_\mu^\mu\rangle_{\rm ren}$, but also 
$\langle\hat T_0^0\rangle_{\rm ren}$ etc.
As the derivation of these quantities is more involved and includes
other contributions (e.g., $\langle\hat T_{01}^2\rangle_{\rm ren}$)
as well as a time-integration, it is not clear that the entanglement 
between $\hat T_{\mu\nu}$ and the corresponding operator $\widehat R$
can be omitted, for example  
\bea
\langle\hat T_{\mu\nu}\widehat R\rangle_{\rm ren}
\stackrel{?}{\approx}
\langle\hat T_{\mu\nu}\rangle_{\rm ren}
\langle\widehat R\rangle_{\rm ren}
\,.
\ea

In summary, the semi-classical theory suggests its own demise by
predicting a trajectory $R(t)$ which is not continuously
differentiable ($\ddot R$ diverges) and hence a curvature singularity.
As a result the fate of the circle after the bounce remains unclear in
the sense that it cannot be determined within the semi-classical
picture without additional arguments.
Assuming that the circle re-enters the region of validity of the
semi-classical theory after the bounce, there are two possibilities.
After jumping from one ridge to the other, the trajectory could turn
around and continue towards the normal regime in the middle again
($\dot R$ decreases). 
In this case, the circle would perform an eternal oscillation.
Alternatively, the trajectory could turn to the edge and enter the
anomalous regime.
In this situation, the velocity would increase forever and
asymptotically reach the speed of light.
These run-away solutions are somewhat similar to a trace anomaly
induced inflation in 3+1 dimensional gravity.
Interestingly, the trace anomaly of both, a scalar field in 1+1
dimensions on the one hand and that of, say, a vector field in 3+1
dimensional gravity on the other hand are such that they prevent the
collapse to a metric singularity $R=0$ and admit run-away solutions.
(The Casimir effect is not substantial in that respect, 
cf.~Fig.~\ref{no-cas}.)

\section{Outlook}

As it became evident from the previous discussions, a more exhaustive
analysis would require the investigation of the full quantum
back-reaction.
At a first glance, one could expect that this is provided by quantizing the 
effective action (\ref{Leff}) for the geometric sector, 
cf.~Eq.~(\ref{polyakov}) below, after integrating out the field $\hat\Phi$.
However, it is by no means clear that such an approach is indeed equivalent
to a full quantization of the (geometric) string degrees of freedom 
interacting with the quantum field.
In view of the non-trivial relation between the canonically conjugated
momentum $P$ and the velocity $\dot R$ as well as the occurrence of anomalies 
and the associated non-linearities, a rigorous treatment is apparently 
rather involved.
Furthermore, an {\em ab initio} quantization has to account for non-spherical
geometries, i.e., deviations from the rotational symmetry, since the monopole 
mode inherently couples to higher multipole modes and their 
quantum 
fluctuations -- in contrast to the semi-classical case considered above.

\section{Negative Energies and Stability}

One might ask whether the breakdown of the semiclassical treatment for
the toy model under consideration is just an artifact caused by the
special features of this 1+1 dimensional model or whether it
represents a generic property also for 3+1 dimensional gravity 
(plus quantum fields). 
Even though this question cannot be addressed completely without
solving the full (3+1 dimensional) scenario, one can compare
characteristic features of both systems
(at the semiclassical level). 

For the classical string (without quantum fields) as described by the 
Nambu-Goto action, the (conserved) energy is positive 
$E=\pi\sigma R/\sqrt{1-\dot R^2}$.
Nevertheless, the equation of motion admits singular solutions
(collapse to $R=0$) in close analogy to classical gravity 
(cf.~the singularity theorems).

Owing to the trace anomaly and the Casimir effect, the renormalized
energy of the quantum field (in the background of the string) is
negative in the vacuum state. 
However, that does not imply that the vacuum state of the quantum
field itself (in the semiclassical treatment) is unstable.
After a normal mode expansion
\bea
\hat\Phi(\tau,\varphi)
=
\sum\limits_{m=0}^\infty
\hat Q_m^{(c)}(\tau)\cos(m\varphi)+
\sum\limits_{m=1}^\infty
\hat Q_m^{(s)}(\tau)\sin(m\varphi)
\,,\!\!\!\!\!\!\!\!
\nn
\ea
the Hamiltonian decouples and is positive for each mode
\bea
\hat H_\Phi^\tau=\sum\limits_{m=0}^\infty\;\sum\limits_{\xi=(c,s)}
\left(
\hat P_{m,\xi}^2+
m^2\hat Q_{m,\xi}^2
\right)
\,.
\ea
Apart from the zero mode $m=0$, all field modes possess a unique
ground state -- but even for the zero mode, the quantum evolution is  
not unstable.
In summary, the dynamics of the quantum field itself [in the
semiclassical treatment, i.e., for a fixed trajectory $R(\tau)$] 
does not display any instabilities. 
The stability of the full theory (string plus field), however, 
lies outside the scope of the present investigations 
(i.e., is still unclear) in view of the breakdown of the semiclassical
treatment. 

The negative (renormalized) energy of the quantum field cannot be
attributed to single modes. Instead, being similar to vacuum polarization
effects, it is a consequence of the renormalization procedure.  
This is a generic feature of quantum fields in curved space-times and
occurs in the 3+1 dimensional scenario as well.

In view of the above similarities, one might conjecture that the
results found for the 1+1 dimensional toy model -- i.e., the breakdown  
of the semiclassical treatment and a possible bounce caused by the
trace anomaly --
are not just artifacts of this toy model but shed some light on the 3+1
dimensional situation as well.

\section*{Appendix A: Renormalization}

Starting form the obvious condition that the (renormalized)
energy-momentum tensor of the total system (quantum field plus string)
is conserved with respect to the (flat) 2+1 dimensional embedding
space-time   
\bea
\partial_A T^{AB}_{\rm total}=0
\,
\ea
one can show the compatibility of the two conditions
\bea
\nabla_\mu\langle\hat T^\mu_\nu\rangle_{\rm ren}=0
\;\leftrightarrow\;
\sigma_{\rm ren}={\rm const}
\,,
\ea
for the renormalized energy-momentum tensor of the quantum field and
the string tension, respectively. 

As a consequence of Eq.~(38), the argument works in both directions:
Demanding $\nabla_\mu\langle\hat T^\mu_\nu\rangle_{\rm ren}=0$ 
in order to ensure energy conservation of the quantum field 
separately in the presence of a Killing vector $\xi$ of the induced
metric, one deduces $\sigma_{\rm ren}={\rm const}$.
Conversely, one may start with the assumption 
$\sigma_{\rm ren}={\rm const}$ and derive
$\nabla_\mu\langle\hat T^\mu_\nu\rangle_{\rm ren}=0$ from
$\partial_A T^{AB}_{\rm total}=0$.

Since the decomposition of the total energy-momentum tensor 
$T^{AB}_{\rm total}$ into the contributions of the quantum field and the
string is not unique, one can even impose alternative renormalization 
conditions. 
When renormalizing $\langle\hat T^\mu_\nu\rangle$, its divergent part
has to be absorbed into $\sigma$, while the finite part of the
counter-term remains undetermined. 
Here  we have employed a  ''minimal subtraction scheme'', 
cf.~Eqs.~(\ref{reg-T}) and (\ref{reg-sigma}).
As an alternative renormalization scheme, for example, one may impose the
condition  that the trace still vanishes 
$\langle\hat T^\mu_\mu\rangle_{\rm ren}=0$ and abandon the property
$\nabla_\mu\langle\hat T^\mu_\nu\rangle_{\rm ren}=0$ instead.
Such a renormalization procedure could be envisaged as a
''non-minimal subtraction scheme''.
Instead of employing Eq.~(\ref{reg-T}) we could split up 
$\langle\hat T_{\mu\nu}\rangle_{\rm bare}$ via
\bea
\label{reg-T-alt}
\langle\hat T_{\mu\nu}\rangle_{\rm bare}
=
\frac{1}{\epsilon}\,{g}_{\mu\nu}
+
\frac12
{\mathfrak C}_{\rm tr}\,{\cal R}
\,{g}_{\mu\nu}
+
\widetilde{\langle\hat T_{\mu\nu}\rangle}_{\rm ren}
\,,
\ea
with
\bea
\widetilde{\langle\hat T_\mu^\mu\rangle}_{\rm ren}=0
\;\leadsto\;
\nabla_\mu\widetilde{\langle\hat T^\mu_\nu\rangle}_{\rm ren}\neq0
\,,
\ea
and, consequently,
\bea
\label{reg-sigma-alt}
\sigma_{\rm bare}
=
-\frac{2}{\epsilon}
-{\mathfrak C}_{\rm tr}\,{\cal R}
+\widetilde{\sigma}_{\rm ren}
\,.
\ea
Within this alternative renormalization scheme 
$\widetilde{\sigma}_{\rm ren}$ depends on the geometry
\bea
\widetilde{\sigma}_{\rm ren}=
\widetilde{\sigma}_{\rm ren}(R,\dot R,\ddot R)
\,,
\ea
while the equation of motion for $R(t)$ and the total energy remain 
the same within both schemes. 
Such a non-minimal subtraction scheme employed for renormalization of
the of the quantum-field sector provides the possibility of inducing a
space-time dependence of characteristic parameters 
(here simply the string tension) of the classical (geometric) sector.

\section*{Appendix B: Effective action}

Instead of using the equations of motion, one could investigate the
back-reaction by means of the effective action, cf.~\cite{nagatani}
and also \cite{brevik}. 
In 1+1 dimensions, the trace anomaly can be deduced from the
Liouville-Polyakov effective action \cite{polyakov}
\bea
\label{polyakov}
{\mathfrak A}_{\rm eff}=\frac{1}{96\pi}\int d^2x\,\sqrt{-{g}}\,
{\cal R}\,\Box^{-1}\,{\cal R}
\,.
\ea
In view of the inverse of the d'Alembert operator $\Box^{-1}$, this
expression appears to be non-local, but inserting the value for the
Ricci scalar in terms of the conformal coordinates, for example, it
turns out that it is in fact, effectively local (no particle creation), 
and agrees with the term proportional to ${\mathfrak C}_{\rm tr}$
in the effective Lagrangian in Eq.~(\ref{Leff}) after a coordinate
transformation to the laboratory time.  

However, in this na{\"\i}ve treatment the Casimir term is missing and  
thus requires additional considerations.
This already illustrates a potential danger of the effective action
method -- the fact the total effective Lagrangian is just as a sum of
the terms proportional to ${\mathfrak C}_{\rm tr}$ and the (static)
Casimir effect (in addition to the classical terms) is not 
{\em a priori} obvious.

In order to compare our calculation with the results of
Ref.~\cite{nagatani}, a few remarks are in order:
\begin{itemize}
\item Switching to the Euklidean signature -- as done in
Ref.~\cite{nagatani} -- is a well-defined procedure for the
calculation of the static Casimir energy, but not for general
time-dependent systems.
Hence the derivation of the main result as a sum of the two
contributions is somewhat {\em ad hoc}. 
\item Contrary to the claims in Ref.~\cite{nagatani}, the scenario
under consideration does not exhibit the dynamical Casimir effect 
(in the sense of particle creation) due to the conformal invariance.
\item Furthermore, Ref.~\cite{nagatani} uses a non-relativistic
kinetic term (introduced by hand). 
Therefore, the results can be compared with ours in the adiabatic
regime $\dot R\ll1$ only. 
\end{itemize}
In addition to the problems mentioned in Ref.~\cite{dowker}, for
example, a conclusive discussion of a possible non-trivial dependence 
of the renormalized string tension $\sigma_{\rm ren}(R,\dot R,\dots)$
in dependence of the renormalization scheme (see Appendix A) by means
of the effective action is not obvious. 

Having in mind the various potential difficulties of the effective
action method indicated above, we would like to emphasize that, 
in contrast to this approach, the calculations based on the
renormalized expectation values of the energy-momentum tensor 
$\langle\hat T_{\mu\nu}\rangle_{\rm ren}$ together with employing the
equations of motion, are apparently less ambiguous and more evident.

\newpage

\section*{Acknowledgments}

G.~P.~and M.~R.~acknowledge support from the DFG.
R.~S.~gratefully acknowledges financial support by the Emmy-Noether
Programme of the German Research Foundation (DFG) under grant 
No.~SCHU 1557/1-1 and by the Humboldt foundation.   

\addcontentsline{toc}{section}{References}


\begin{thebibliography}{99}
\bibitem{example}
As an example, let us consider the semi-classical Einstein equations
(\ref{einstein}) for a quantum state $|\Psi\rangle$ describing a
macroscopic superposition
$|\Psi\rangle=(|\Psi_1\rangle+|\Psi_2\rangle)/\sqrt{2}$.
%
The state $|\Psi_1\rangle$ corresponds to a large mass $M$ at the point
$x_1$ whereas $|\Psi_2\rangle$ describes the same mass $M$ at a 
(macroscopically) different position $x_2$.
%
The semi-classical Einstein equations (\ref{einstein}) predict a
gravitational field 
(since $\langle\Psi_1|\hat T^\mu_\nu|\Psi_2\rangle_{\rm ren}=0$) 
as if there was half the mass $M/2$ at $x_1$ and the other half $M/2$
at $x_2$.  
%
However, in reality this macroscopic superposition $|\Psi\rangle$ will
exhibit an extremely fast decoherence with a negligible energy flux
due to the interaction with the environment leading to a classical
mixture $\hat\varrho=(\hat\varrho_1+\hat\varrho_2)/2$.
%
Hence one would naturally expect a classical mixture of gravitational
fields.
\end{thebibliography}
\end{document}